\documentclass[aps,prl,reprint,groupedaddress]{revtex4-1}

\usepackage{graphicx}
\usepackage{subfigure}
\usepackage{amssymb}
\usepackage{amsmath}
\usepackage{natbib}

\newcommand{\bfu}{\mbox{\boldmath$u$}}

\newcommand{\bfzhat}{\mbox{\boldmath $ {\hat z} $ } }
\newcommand{\bfB}{\mbox{\boldmath$B$}}

\newcommand{\bmnabla}{\mbox{\boldmath$\nabla$}}

\begin{document}

\title{Strong field dynamo action in rapidly rotating convection with no inertia}

\author{David W. Hughes}
\email[]{d.w.hughes@leeds.ac.uk}
\affiliation{Department of Applied Mathematics, University of Leeds, Leeds LS2 9JT, UK}
\author{Fausto Cattaneo}
\affiliation{Department of Astronomy and Astrophysics, University of Chicago, 5640 South Ellis Avenue, Chicago, IL 60637, USA}

\date{\today}

\begin{abstract}
The Earth's magnetic field is generated by dynamo action driven by convection in the outer core. For numerical reasons, inertial and viscous forces play an important role in geodynamo models; however, the primary dynamical balance in the Earth's core is believed to be between buoyancy, Coriolis and magnetic forces. The hope has been that by setting the Ekman number to be as small as computationally feasible, an asymptotic regime would be reached in which the correct force balance is achieved. However, recent analyses of geodynamo models suggest that the desired balance has still not yet been attained. Here we adopt a complementary approach consisting of a model of rapidly rotating convection in which inertial forces are neglected from the outset. Within this framework we are able to construct a new branch of solutions in which the dynamo generates a strong magnetic field that satisfies the expected force balance. The resulting strongly magnetized convection is dramatically different to the corresponding solutions in which the field is weak.
\end{abstract}


\maketitle

Convective fluid motions in the liquid outer core drive the dynamo responsible for maintaining the Earth's magnetic field over geological time scales \citep{RK_2013}. Understanding the nature of the geodynamo is essential in explaining the observed spatial and temporal variations of the geomagnetic field. The physical properties of the core are characterized by strong rotation, very small viscosity and vigorous buoyancy. These translate into an extremely small Ekman number $E$ (the ratio of viscous to Coriolis forces) and a Rayleigh number $Ra$ (a measure of the thermal driving) far in excess of its critical value at the onset of convection. Even though the system is very strongly supercritical, the rotational constraints are such that the convective turnover time greatly exceeds the rotational period; inertial forces are therefore completely negligible. Another distinctive feature of the geodynamo is that, unlike in stellar dynamos, which operate at extremely high values of the magnetic Reynolds number $Rm$ (the ratio of advection to diffusion of magnetic fields), with values  $O(10^{10} - 10^{14})$, in the geodynamo $Rm$ is moderate, $O(10^2 - 10^3)$. These moderate values, based on adopting the outer core radius as the length scale, are only somewhat larger than those required for the onset of dynamo action; they therefore impose a very strong constraint on the characteristic scales of the motion. In particular, if the scales were controlled by viscous stresses, then the horizontal extent of a convective cell would be $O \left( E^{-1/3}  d \right)$, where $d$ is some characteristic length scale  \citep[e.g.][]{Chandra_1961}. In the Earth, $E \sim 10^{-15}$; taking $d$ as the depth of the liquid core thus leads to very narrow convective cells, of width $O\left( 10^2 \right) \mathrm{m}$. Given the inferred slow characteristic speed of the motions in the core, $10^{-3} \, \mathrm{m} \mathrm{s}^{-1}$, and magnetic diffusivity $\eta \approx 1 \, \mathrm{m}^2 \mathrm{s}^{-1}$, the \textit{local} magnetic Reynolds number based on such convective cells is $Rm_l=O \left( 10^{-1} \right)$. This small value of $Rm_l$ is problematic. Although dynamo action is not completely ruled out, the only possibility is a mean-field, low $Rm$ dynamo \cite{CS_1972, Soward_1974}. In such a dynamo, the field, however, is weak and hence incompatible with estimates of the Earth's field strength \citep{RS_1992}. Another dynamical ingredient is therefore required to control the scale of the motions. This has led to the idea that the convection in the Earth's core is strongly influenced by the magnetic field; the resulting balance of forces is then between Magnetic, buoyancy (Archimedean) and Coriolis forces, with inertial forces and viscous stresses playing no role at all --- so-called MAC balance.

The notion of MAC balance, which has become central to an understanding of the geodynamo, places a severe constraint on the nature of the dynamo action that must be taking place within the Earth's core. For dynamo saturation to occur, it is of course necessary that the magnetic forces modify the flow. In many cases these changes are subtle and leave the velocity in the saturated state close to what it would have been in the absence of a magnetic field; we may classify these as \textit{weak field dynamos}. This however is not the case if MAC equilibrium is realized. Here the influence of the field is to cause substantial changes in the fluid velocity that allow the field to grow to a large amplitude; we classify this as a \textit{strong field dynamo}. It is worth noting that a very different distinction may be drawn when classifying dynamos --- namely that between small-scale and large-scale dynamos. In this more traditional decomposition, the emphasis is on the scale of the magnetic field generated, in comparison with that of the velocity; by contrast, the distinction between weak and strong field dynamos addresses the dominant force balance. Thus it is possible to have weak and strong field dynamos of both the large- and small-scale variety. 

Even though there are powerful theoretical arguments supporting the existence of a strong field dynamo, constructing one has proven to be a non-trivial exercise. Traditionally, the geodynamo problem has been tackled through the computational framework of MHD in spherical geometry. There are, broadly speaking, two complementary approaches. The majority of computations have progressed through decreasing Ekman numbers, with the idea that the regime of MAC balance will emerge naturally. That said, to give a sense of the magnitude of the difficulty in attaining the regime of the Earth, we note that even the most heroic computations to date can resolve Ekman numbers no lower than $10^{-7}$. Of course, the hope is that an Earth-like solution will emerge without having to close this considerable gap. This ongoing project, has, over the past two decades, provided tremendous insight into dynamo modeling \citep[e.g.][]{GR_1996, KB_1997, COG_1998, KMS_2008, SR_2009, NS_2015}. However, crucially, simulations to date typically do not exhibit MAC balance. From an analysis of scaling laws applied to a large number of numerical dynamo models, \cite{KB_2013} determine that current models are compatible with a scaling in which viscosity plays a prominent role, with no support for a MAC balance. A similar conclusion is reached by \cite{SKA_2012}, who conclude that in most of the models considered, the characteristics of convection are changed only very slightly by the growth and saturation of a dynamo-generated magnetic field, i.e.\ that the dynamos are weak field dynamos. It is though of interest to note that in the plane layer dynamo simulations of \cite{SH_2004}, there is a transition to large-scale convection at low Ekman number. Indeed, the horizontal convection scale grows until it is comparable with the size of the domain, which, for computational reasons was constrained to be fairly small. Possibly because of this restriction, the large-scale convective flow is not capable of maintaining the strong field indefinitely; the field decays and the flow reverts to being small-scale. Field amplification can then restart, leading eventually to another transition to large-scale convection. Thus the amplitude of the generated magnetic field fluctuates strongly in time.

The alternative approach to retaining all of the terms in the momentum equation and striving for the lowest possible value of $E$, is to neglect inertia from the outset \citep[e.g.][]{GR_1995a, GR_1995b}. Our aim in this letter is to investigate such a model in order to tackle the fundamental problem of the existence of dynamo solutions in MAC balance. As we shall see, inertia-free systems not only have compelling physical motivation in terms of the geodynamo, but also afford a means of convincingly identifying MAC balance if and when it arises.

The neglect of inertia in geodynamo modeling comes not from the fact that the fluid Reynolds number $Re$ is small, but that the rotation is strong. Indeed, in the Earth, $Re$ is huge, rendering the problem extremely computationally demanding. In order to make best use of computational resources, we therefore study the simplest possible model in which inertial forces can be neglected in a self-consistent manner. Following the pioneering study \cite{JR_2000}, we consider rapidly rotating, Boussinesq convection at infinite Prandtl number (the ratio of viscous to thermal diffusivity) in a plane layer. This allows us to study inertia-less, turbulent dynamo action over an extended spatial domain. Under the simplifying assumption that the rotation axis is aligned with gravity (along the $z$-axis), the dimensionless governing equations may be written as
\begin{equation}
\bfzhat \times \bfu = - \bmnabla p + \left( \bmnabla \times \bfB \right)  \times \bfB + R \theta \bfzhat + E \nabla^2 \bfu ,
\label{eq:mom}
\end{equation}
\begin{equation}
\bmnabla \cdot \bfu = 0,
\end{equation}
\begin{equation}
\frac{\partial \bfB}{\partial t} = \bmnabla \times (\bfu \times \bfB) + \frac{1}{q} \nabla^2 \bfB ,
\end{equation}
\begin{equation}
\frac{\partial \theta}{\partial t} + \bfu \cdot \bmnabla \theta = \bfu \cdot \bfzhat + \nabla^2 \theta ,
\label{eq:temp}
\end{equation}
where $\bfu$ is the fluid velocity, $\bfB$ is the magnetic field, and $\theta$ is the temperature perturbation. Equations~(\ref{eq:mom})--(\ref{eq:temp}) are obtained by scaling lengths with the layer depth $d$, and times with $d^2/\kappa$, where  $\kappa$ is the thermal diffusivity. The dimensionless quantities $R$, $E$ and $q$ are the rotational Rayleigh number, Ekman number and Roberts number respectively, defined by
\begin{equation}
R = \frac{g \alpha \beta d^2}{2 \varOmega \kappa} , \quad E = \frac{\nu}{2 \varOmega d^2}, \quad q = \frac{\kappa}{\eta} ,
\end{equation}
where $g$ is the (constant) acceleration due to gravity, $\alpha$ is the coefficient of thermal expansion, $\beta$ is the temperature gradient of the basic state, $\varOmega$ is the rotation rate, and $\nu$ is the kinematic viscosity. 

The distinctive feature of the model is that only the Coriolis acceleration appears on the left hand side of (\ref{eq:mom}), which is now a diagnostic equation. Equation~\eqref{eq:mom} is obtained by considering the momentum equation, including inertial terms, and then letting the Prandtl number $Pr = \nu /\kappa$ tend to infinity, \textit{whilst keeping both $R$ and $E$ finite}. Thus this limit should \textit{not} be thought of as either $\nu \to \infty$ or $\kappa \to 0$ with other parameters kept fixed. A crucial aspect of taking $Pr \to \infty$ is that the magnetic field must scale with $Pr^{1/2}$ if it is to play any role.  As such, a comparison of the magnitudes of the magnetic and kinetic energies  in this limit becomes meaningless; in particular, it cannot be used to distinguish between weak and strong field dynamos. As discussed above, given the motivation of the problem, it is thus most natural to make this distinction by reference to the extent to which the convection is influenced by the magnetic field.

We solve equations~(\ref{eq:mom})--(\ref{eq:temp}) in a domain with square horizontal cross-section of length $5 d$. All variables are periodic in the horizontal directions; the lower and upper boundaries are impermeable, stress-free and perfectly conducting, both thermally and electrically. The equations are solved numerically by standard fully de-aliased pseudospectral methods \citep{CEW_2003}, with a resolution of $256 \times 256 \times 97$ or $512 \times 512 \times 193$.

We are motivated by the case of small Ekman number~$E$. 
Near onset, the convection assumes a columnar structure, with a more complicated sheet-like pattern emerging as $R$ is increased. In this hydrodynamic state, the balance is between the Coriolis force, buoyancy and viscous stresses; simple considerations of the vorticity equation show that the characteristic horizontal scale $\ell \sim E^{1/3}$. The expectation is that for large enough values of $q$ (i.e.\ large enough $Rm$), these flows will act to generate magnetic field. The crucial question is then what sort of field emerges. As mentioned above, one possibility is that the generated field is weak, in the sense that although the Lorentz force cannot be ignored, it does not lead to a significant change in the convective pattern, the dynamical balance remaining essentially the same. The other more interesting possibility is that the Lorentz force drives the system to a new state in which MAC balance is attained. We show presently that, depending on parameters, both types of dynamo are possible.

We have performed an extensive exploration of the nature of the convection and possible dynamo action when rotational effects are important. It is, however, important to recognize that, having neglected inertial terms from the outset, we are not reliant on pushing $E$ down to extremely low values in order to access the appropriate asymptotic regime; the problem can thus be addressed via values of $E$ that are small, but not so small as to be computationally overwhelming. For each value of $E$ we vary $R$ to study convection from close to onset up to substantially supercritical; for each case we vary $q$ to explore dynamo action at different values of $Rm$. Each calculation is first evolved from random initial conditions to an unmagnetized convecting state. For infinite $Pr$, the fluid motions have little variation with depth; depending on the parameter values, the convective pattern takes the form of either cylindrical columns or sheets. A small magnetic perturbation is then introduced. For those cases where dynamo action ensues, the magnetic energy first grows exponentially (the kinematic phase), before reaching an amplitude at which Lorentz forces modify the flow, the field growth ceases, and the system settles down to a stationary state. Our interest here is in the nature of these dynamical MHD states. We present results from two representative cases, with $E$, $R$ and $q$ given by (i)~ $10^{-3}$, $500$, $5$ and (ii)~$10^{-4}$, $500$, $20$ respectively. We assess the influence of rotation by a convective Rossby number, $R_c$, defined as the ratio of the rotation period to the convective turnover time, with the latter being determined \textit{a posteriori}. For these two cases, $R_c$ is roughly $20$ times smaller in case~(ii) than case~(i).

Several features of the solutions to the governing equations~(\ref{eq:mom})--(\ref{eq:temp}) are illustrated in figure~1. In case~(i), the growth and saturation of magnetic energy is accompanied by a slight decrease in kinetic energy. (Although, as noted earlier, a comparison of the levels of magnetic and kinetic energy is not meaningful, comparison of their temporal evolution does provide useful information.) Even though the velocity field has to be different in the kinematic and dynamic phases, the changes are subtle and are certainly not reflected in any dramatic changes to the convective planform; in particular, the horizontal scale of the convection remains the same. In case~(ii), the growth of the magnetic energy is again reflected in changes to the kinetic energy, although here the latter increases. The dynamical effect of the magnetic field is thus to make the convection more vigorous. This is manifest in the dramatic changes to the convective planforms between the kinematic and dynamic regimes. In the kinematic regime, the horizontal scales are much smaller than in case~(i), reflecting the smaller value of $E$. Strikingly, the dynamic regime shows the emergence of a much larger scale pattern of convection.

\begin{figure}
       \includegraphics[clip=true,width=9cm]{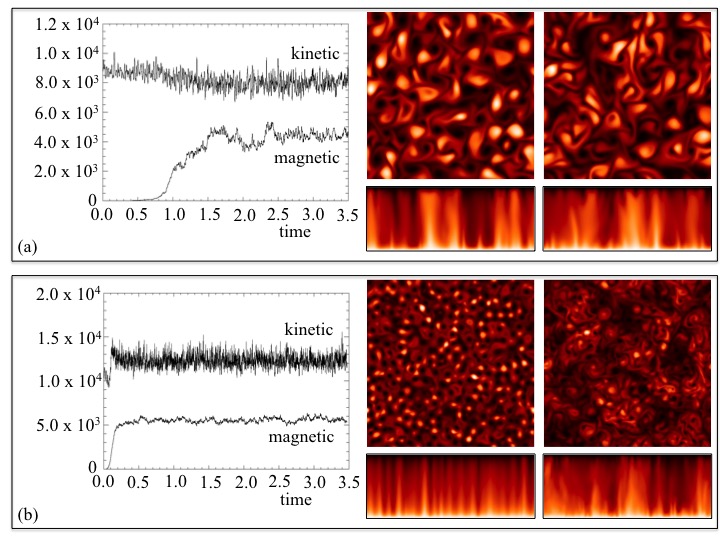}
       \caption{\label{fig:fig_01} Nonlinear evolution of convectively driven dynamos for (a) case~(i), and (b) case~(ii). The leftmost panels show the time histories of kinetic and magnetic energy densities. The other panels are density plots of the temperature at representative times in the kinematic (left) and dynamic (right) regimes. The square panels are horizontal ($xy$) slices near the upper boundary; the rectangular panels are vertical $xz$ slices at $y=0$. Light and dark tones correspond to hot and cold fluid. Case~(i): $E=10^{-3}$, $R=500$, $q=5$. Case~(ii): $E=10^{-4}$, $R=500$, $q=20$. }
\end{figure}

These ideas can be made more evident by exploiting an interesting consequence of the absence of inertia. The momentum equation~\eqref{eq:mom} is linear in $\bfu$, thus allowing the velocity to be decomposed into the sum of a thermal component $\bfu_T$, satisfying
\begin{equation}
\bfzhat \times \bfu_T = - \nabla p_T + R \theta \bfzhat + E \nabla^2 \bfu_T,
\qquad 
\nabla \cdot \bfu_T = 0,
\end{equation}
and a magnetic component $\bfu_M$ defined by
\begin{equation}
\bfu_M = \bfu - \bfu_T.
\end{equation}
This decomposition provides a useful way to visualize the relevant contributions of thermal and magnetic forces to the structure of the flow, as shown in figure~2. In case~(i), the total velocity is almost identical to the thermal velocity, with just a small change resulting from the magnetic velocity, which is needed to saturate the dynamo. The energy in $\bfu_M$ is, on average, $13\%$ of that in $\bfu_T$. By contrast, in case~(ii), the influence of the thermal and magnetic velocities is comparable; features in each are reflected in the total velocity. In this case the energy in $\bfu_M$ is $50\%$ of that in $\bfu_T$. These considerations strongly suggest that case~(i) is a weak field dynamo, whereas case~(ii) is a strong field dynamo.
\begin{figure}
       \includegraphics[clip=true,width=9cm]{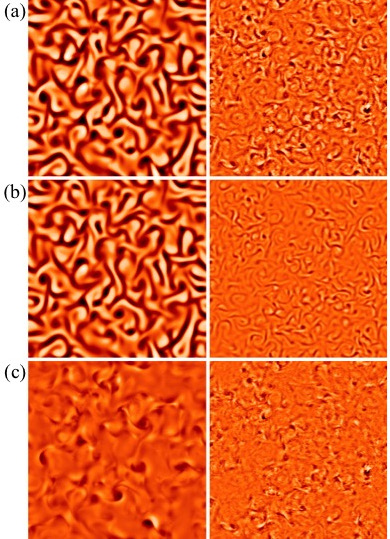}
       \caption{\label{fig:fig_02} Snapshots of the flow structure in the dynamical regime. Density plots of the vertical components of (a) the full velocity ($w$), (b) the thermal component ($w_T$), and (c) the magnetic component ($w_M$), near the upper boundary, for the weak field dynamo (case~(i), left column) and strong field dynamo (case~(ii), right column). Light and dark tones correspond to up-flowing and down-flowing material. The  weak and strong field cases are scaled independently, but the same color scaling is adopted for all three plots within the two cases.}
\end{figure}

We recall that the idea of MAC balance was introduced to argue that dynamo solutions could be found whose characteristic scales were not determined by viscosity. Clearly this cannot be the case for a weak field solution; inspection of equation~\eqref{eq:mom} shows that if the Lorentz force is not a factor in determining the scale then this role has to be played by the viscous forces.  It remains to be seen whether, in strong field solutions, the viscous terms are still significant, i.e.\ do strong field solutions necessarily lead to MAC balance? This point can be further explored by considering the power spectra of the solutions, as shown in figure~3. In case~(i), the kinetic energy spectra for both the kinematic and dynamic regimes peak at approximately the same scale. This is consistent with the designation as a weak field dynamo, namely one in which the Lorentz force is unable to change the overall structure of the flow. By contrast, in case~(ii), the characteristic scale apparent in the kinematic spectrum is completely lost in the dynamic regime. Indeed, the shift of energy to larger scales suggests that viscosity no longer plays a role in determining the flow structure, which we interpret as evidence that MAC balance is being attained. This however should not be interpreted as implying that there are no small scales; indeed, inspection of the spectrum shows that the velocity in the dynamical regime has more energy at small scales than in the kinematic regime.
\begin{figure}
       \includegraphics[clip=true,width=9cm]{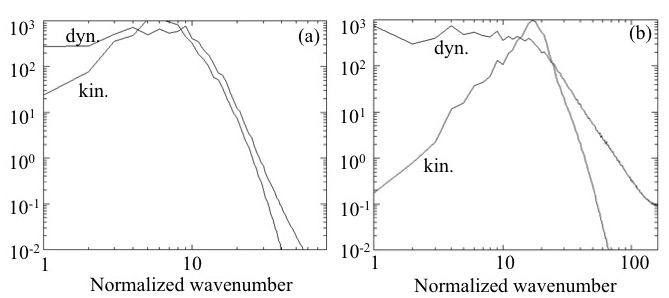}
       \caption{\label{fig:fig_03} Transverse kinetic energy spectra for the kinematic and dynamical regimes for (a) the weak field and (b) the strong field dynamo. The individual spectra are computed from two-dimensional horizontal Fourier transforms; the spectra are then averaged over the convective interior ($0.1 < z < 0.9$) and in time. The normalized wavenumber is the horizontal wavenumber $k_h = \sqrt{k_x^2 + k_y^2}$ in units of $2 \pi/5$.}
\end{figure}

It is important to speculate why the different choice of parameters results in such different behavior. In answering this question we note that the effect of a magnetic field, whether self-generated or externally imposed, may be either to act as a constraint or, alternatively, to act to alleviate another constraint. For example, in standard magnetoconvection, the presence of an imposed uniform magnetic field is always stabilizing \citep[see][]{WP_2014}; however, if the system is strongly rotating then there are circumstances in which the presence of a uniform field actually facilitates the development of convection \citep{Chandra_1961, ER_1970}. The system here is a little more complicated since the magnetic field is self-generated, rather than imposed; nonetheless, it is reasonable to assume that some of the same principles apply. Thus we believe that the MAC equilibrium is observed only in case~(ii) since it is $20$ times more rotationally constrained than case~(i). Given this, we may conjecture that  the existence of strong field solutions requires two conditions: strong rotation and a sufficiently high $Rm$ should motions develop. These conditions are likely to be satisfied, for instance, in some neighborhood of marginal hydrodynamic stability.


How do these considerations relate to the modeling of the geodynamo? On the one hand, the parameters we have considered are not Earth-like; indeed, the Ekman number is not particularly small and the Prandtl number is formally infinite. That said, our strong field solutions have the dynamic balance that is believed to be crucial to the geodynamo. This should be contrasted with the alternative approach, in which one tries to match the realistic physical parameters as closely as possible, but which offers no guarantee of exploring the physically relevant branch of solutions. We note here the recent work of \cite{Dormy_2015}, which shows that simply pushing the parameters towards their values in the Earth, at ever-increasing computational cost, does not lead to more Earth-like solutions.

It is interesting to note that in one of the pioneering computations that set in motion the modern approach to modeling the geodynamo \citep{GR_1995a,GR_1995b}, inertial terms were in fact neglected. In spite of the computational limitations of the time, the solutions were remarkable in displaying convectively driven dynamo action with a reversal. However, it is not obvious whether the solutions were in MAC balance; with the limitations in computational resolution, the small scales that may be important in determining MAC balance had to be suppressed by hyperdiffusivities. In view of the results of this article, it would therefore be of great interest to revisit the inertia-less spherical shell problem with today's computational resources.

\begin{acknowledgments}
We are very grateful to Chris Jones for many helpful discussions and suggestions. This work was supported by the Natural Environment Research Council under grant NE/J007080/1 and by the National Science Foundation sponsored Center for Magnetic Self Organization at the University of Chicago. Computations were undertaken on ARC1 and ARC2, part of the High Performance Computing facilities at the University of Leeds. This work also used the DiRAC Data Centric system at Durham University, operated by the Institute for Computational Cosmology on behalf of the STFC DiRAC HPC Facility (www.dirac.ac.uk). This equipment was funded by BIS National E-infrastructure capital grant ST/K00042X/1, STFC capital grants ST/H008519/1 and ST/K00087X/1, STFC DiRAC Operations grant  ST/K003267/1 and Durham University. DiRAC is part of the National E-Infrastructure.
\end{acknowledgments}

\bibliography{refs}

\end{document}